\documentclass[sigconf]{acmart}

\usepackage{booktabs} 

\title{
Positivity Bias in Customer Satisfaction Ratings}

\def\pprw{8.5in}
\def\pprh{11in}

\begin{document}

\copyrightyear{2018}
\acmYear{2018}
\setcopyright{iw3c2w3}
\acmConference[WWW '18 Companion]{The 2018 Web Conference Companion}{April
23--27, 2018}{Lyon, France}
\acmBooktitle{WWW '18 Companion: The 2018 Web Conference Companion, April
23--27, 2018, Lyon, France}
\acmPrice{}
\acmDOI{10.1145/3184558.3186579}
\acmISBN{978-1-4503-5640-4/18/04}

\fancyhead{}

\author{Kunwoo Park}
\affiliation{
  \institution{KAIST}
  \city{Daejeon} 
  \country{South Korea} 
}
\email{kw.park@kaist.ac.kr}

\author{Meeyoung Cha}
\affiliation{
  \institution{KAIST}
  \city{Daejeon} 
  \country{South Korea} 
}
\email{meeyoungcha@kaist.ac.kr}

\author{Eunhee Rhim}
\affiliation{
  \institution{Samsung Electronics}
  \city{Suwon} 
  \country{South Korea} 
}
\email{eunhee.rhim@samsung.com}

\begin{abstract}
Customer ratings are valuable sources to understand their satisfaction and are critical for designing better customer experiences and recommendations. The majority of customers, however, do not respond to rating surveys, which makes the result less representative. To understand overall satisfaction, this paper aims to investigate how likely customers without responses had satisfactory experiences compared to those respondents. To infer customer satisfaction of such unlabeled sessions, we propose models using recurrent neural networks (RNNs) that learn continuous representations of unstructured text conversation. By analyzing online chat logs of over 170,000 sessions from Samsung's customer service department, we make a novel finding that while labeled sessions contributed by a small fraction of customers received overwhelmingly positive reviews, the majority of unlabeled sessions would have received lower ratings by customers. The data analytics presented in this paper not only have practical implications for helping detect dissatisfied customers on live chat services but also make theoretical contributions on discovering the level of biases in online rating platforms.
\end{abstract}

\maketitle

\section{Introduction}

An increasing number of products and services ask for customer ratings. Customers are prompted to give feedback after a visit to a bank or hotel, an Uber ride, and more. Websites like Yelp, TripAdvisor, and Angie's List ask consumers to voluntarily rate hundreds of millions of restaurants, shops, and entertainment hotspots around the world. Ratings are invaluable not only for improving customer perception but also for compiling recommendations~\cite{elkahky2015multi,tang2016empirical} and influencing future purchases~\cite{kumar2006research}. More than 90\% of people say they look up online reviews prior to purchases and over 88\% of them trust online reviews as much as personal recommendations~\cite{nielsen}. In fact, a study conducted by Nielsen reports that online reviews are trusted even more than editorialized advertisements that appear on brand websites, television channels, and magazines. These trends indicate that online ratings have become one of the most trusted information sources in e-commerce decisions.

Numerical star ratings in customer reviews (usually ranging from one to five stars) are known to have a ``J-shaped distribution'', where ratings tend to be disproportionately positive. A consumer is more likely to give positive ratings (e.g., 4-5 stars) than negative or moderate ratings (e.g., 1-3 stars); therefore, the average rating is biased toward positive scores. Past research has found that customer ratings may be systematically biased for several reasons. First is the acquisition-led selection bias, where ratings tend to be more positive than the ground truth because they are from purchasers, who are likely positively predisposed~\cite{bikhchandani1992theory}. Second is the social influence bias, where new raters are influenced by existing ratings and, thereby, existing positive ratings dramatically affect future ratings~\cite{salganik2006experimental}. Third is the under-reporting bias, where consumers who are greatly satisfied or dissatisfied are more likely to report a rating~\cite{hu2009online}. This last bias may be amplified when consumers view star rating systems as reflecting attitude extremity or deviation from the midpoint of an attitude scale~\cite{krosnick1993attitude}. The under-reporting bias commonly occurs as a form of \textit{positivity bias}, where positive feedback is more likely to be prevalent in the overall rating system. Understanding such bias is important because it can reveal the latent sentiment and true demand towards services, which is critical for untapping the full business potential. 

In this research, we are given a unique opportunity to study inherent biases in online consumer ratings by having access to a large chat data set from Samsung's customer service department. The live chat system logs text-form chat messages between customers and service agents to provide remote support for various products, including cellphones and televisions. At the end of each live chat session, customers are prompted with a 5-star rating survey asking how satisfied they are with the experience. Respondents, on average, gave positive feedback, with 68.1\% of ratings being 4 or 5 stars. This survey, however, had been answered by only 16.2\% of the chat customers. The remaining 83.8\% left the chat service without giving feedback. Our goal is to infer the missing satisfaction scores for these non-respondents and thereby to understand the true customer ratings of the entire system. Utilizing the session logs for both satisfying and unsatisfying sessions, we propose a deep learning model that efficiently handles chat sequence data. 

Our methodology to predict the latent satisfaction scores from large conversation data is timely, because live chats are becoming a popular channel of customer service (e.g., WeChat's business profiles, Facebook's M, and Skype's helpdesk). They are a critical business operation, offering a direct line of communication with customers. Mining these data is advantageous for several reasons. First, live chats are stylized so that their objectives are focused and limited. Second, they are abundant, providing ample test cases for training. Third, chats are contained within an online environment; therefore, their logs capture all verbal and non-verbal emotional cues. These characteristics make the live chat data suitable for machine-learning tasks.

While many learning algorithms exist, deep neural networks are used in this paper for their advantage in handling sequence data. A customer's mood develops throughout the conversation, which affects one's linguistic choices and interaction frequencies over time (e.g., long pauses, short responses, or apathetic attitude). Such sequence dependency was effectively modeled with recurrent neural networks.

Our paper makes the following key findings. 
\begin{enumerate}

\item We test the positivity hypothesis in the context of customer ratings: {labeled live chat sessions are likely to receive more positive ratings than unlabeled sessions}. We find that the mean satisfaction score of raters is higher (79.7\% positive or neutral) than the inferred satisfaction score for non-raters (45.5\% positive or neutral).  
\item The prediction of ratings was efficiently modeled with the long short-term memory (LSTM)-based neural network. By incorporating non-textual features with text sequences from chat conversations, the proposed LSTM network outperforms the existing feature-based approaches on predicting customer satisfaction.
\end{enumerate}

Consumer ratings are a scarce resource as they require time and effort to collect. While these ratings are critical for a wide variety of applications, in reality most consumers remain silent. Hence, the design and implementation presented in this paper, which was evaluated from an active customer center, have practical industry implications. Our deep learning model can be applied to inferring missing consumer ratings of live chat services in general service domains including electronics, travel booking, and online shopping. This will help us better understand the unbiased ratings of customers, which are fundamental for customer relationship management. 

\section{Theoretical Grounds}
 
Conversation is a major mode of social interaction and occurs on diverse media on the web (e.g., e-mail, social media, wikipedia).

One type of conversation that occurs frequently is between customers and businesses. Proprietary call center logs, after speech-to-text translation, have been studied extensively to improve service designs. One of the first data-driven studies classified whether a call would be resolved quickly or take a long time~\cite{tan2000textual}. Another study classified dialogs into predefined types based on language features such as opening, question, answer, thanks, and closing~\cite{kim2010classifying}. Other study built support vector machine (SVM) classifiers to identify customer intentions and sentiments~\cite{li2016customer}. As more companies adopt online live chat systems over call centers, the comprehensive and complete logs embedded in chats help businesses and researchers better understand customers' needs than logs of conventional telephone-based alternatives.

In particular, by analyzing live chat logs along with customer ratings, one can directly identify factors that contribute to customer satisfaction. This information, in turn, can be used for real-time monitoring of ongoing chats or for evaluating sessions with missing customer ratings. A handful of studies have made efforts in estimating customer satisfaction in online dialogs. Past research proposed machine-learning methods to classify dissatisfied customers by extracting features from chat texts~\cite{park2015mining}. They have found that a random forest model trained on linguistic features such as positive and negative emotions outperforms algorithms trained on other session meta-data such as session length and word count. 
Another research group examined business conversations on Twitter and similarly found affective features drawn from the text to be critical in predicting customer satisfaction~\cite{herzig2016predicting}. Their work finds that personality traits and emotion expressions improve prediction of customer satisfaction when added to more typical text-based features. Based on prior findings that affective expressions are a key determinant of customer satisfaction, this research considers chat text as a primary input.

For the prediction task, deep learning methods are proposed in this research. In particular, the type of deep neural network that we employ is recurrent neural network (RNN), in which connections between units form a cycle~\cite{mikolov2010recurrent}. Unlike feedforward neural networks, which assumes each input is independent, RNNs model dependency between inputs through the cycle.
Its recurrent structure allows it to model sequential information such as time series~\cite{madetecting} and text sequences~\cite{tai2015improved}. Text sequences such as chat logs, for example, contain dependency between words, and thus a word occurrence is meaningful when the model jointly considers which other words preceded. RNNs can handle such text sequences effectively. Recently, deep neural networks have successfully shown their capabilities to model complex relationships on the web. One study proposed a deep query understanding model from text-based personal queries to clicked photos~\cite{jiang2017delving}. Another research group presented a deep memory network to identify attitudes of people~\cite{li2017deep}. In both works, the deep-learning-based approaches outperformed feature-based traditional approaches, respectively.

\begin{figure}[ht]
\centering
\includegraphics[width=\linewidth]{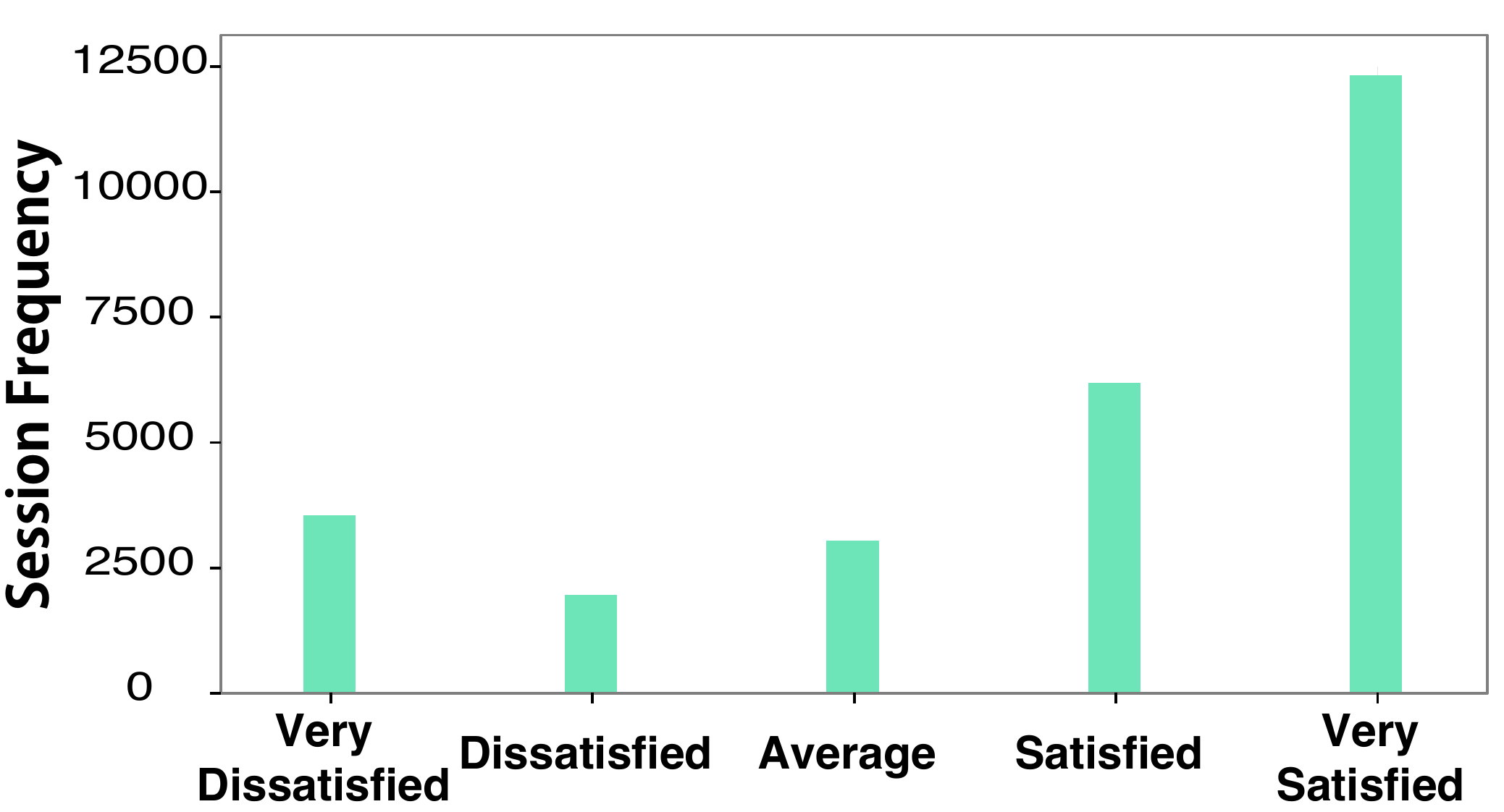}
\caption{Satisfaction score of the live chat data}\label{fig:data_property}
\end{figure}

\section{Data}

\subsection{The live chat system}

Our data set consists of meta-information and text messages of 173,886 chat sessions {in English} and their 5,641,172 speech units between customers and agents from Samsung's live chat service over a one-year period. The chat service is part of the larger customer support operation that runs 24/7 to assist customers who use Samsung products that were sold in the United States. Customers accessing from anywhere in the world can initiate a chat by visiting the web link \url{http://www.samsung.com/us/support/live-chat.html}, upon which they must first type in their names and pick the product category in question, such as dishwashers, digital cameras, computer monitors, printers, or televisions. We chose to study logs in the cell phones category based on criteria of chat frequency, as it contained the highest number of sessions. 

Chat sessions are stylized and their conversation topics are limited to products in service. For instance, sessions start with a greeting message by agents, followed by question-and-answer messages between customers and agents, and finally an ending message by agents. The smallest unit in a chat conversation is called the \textit{utterance}, which is a message that is typed until a speaker hits the enter key, which triggers the current message to be sent to the other party. A single utterance may contain a full sentence. Sometimes it can be part of a sentence, depending on the style of the speaker. Later in the analysis, consequent utterances are consolidated by the level of speaker turn-taking to mitigate individual style differences.

A typical session is terminated by agents thanking customers for using the chat service and asking them to participate in a survey that is prompted after the chat. The survey asks, ``{How would you rate your overall satisfaction with the chat?}'' with the option of ``{Very Dissatisfied}'', to ``{Dissatisfied}'', ``{Average}'', ``{Satisfied}'', and ``{Very Satisfied}''. These survey responses were used as the dependent variable in the prediction task.  

Each session contains a set of meta-information about customers and agents. IP addresses can be used to infer the timezone of each customer. Time stamp information of the chat log, however, is based on the customer center's server time. Hence, prior to analysis, we used the Geolocation API~\cite{ipapi} to obtain the time zone information for each customer and translated the time stamp information into the local time of the corresponding customer. In summary, the following information was gathered in an XML format for each chat session: 
\begin{enumerate}
\item \textit{Customer information:} IP address, geocode, region name, self-identified user name of the customer;
\item \textit{Agent information:} Agent ID and user name;
\item \textit{Chat content:} The list of utterances and their information including the speaker name, time stamp, and the chat text;
\item \textit{Survey result:} The star rating review (from 1-star to 5-star) a customer provided after the chat session;
\item \textit{Session information:} All other session-level data including the chat start time, end time, and disconnecting entity.
\end{enumerate}

Sessions that did not have enough chat content were excluded from analysis. We set this threshold as 4 utterances, considering a typical chat flow (i.e., opening, question, solution, and closing). This led to the elimination of 7,043 sessions that were shorter than 4 utterances, which was 4\% of all sessions. The final data set comprised 166,843 chat sessions.

\subsection{Data properties}

The chat data have several key properties that are important for understanding customers. First, survey results were missing for a large majority of sessions (83.78\%). This is because the survey is participation-based and customers may leave the chat without answering any questions. Figure~\ref{fig:data_property} displays {the} histogram of satisfaction scores from the remaining 16.22\% of the sessions, which follows the expected J-shaped distribution that is commonly seen in online ratings~\cite{hu2009online}. From the survey on satisfaction with chat sessions, the largest proportion of customers (45\%) indicated they were Very Satisfied with the chat experience and a much smaller proportion (14\%) expressed they were Very Dissatisfied with the service. The mean satisfaction score is 3.79 out of 5.0 when based on a score from 1 to 5, and this indicates an overall positive experience. Together, 68.1\% of respondents gave ratings of 4 or 5 stars and 79.7\% gave non-negative ratings (3-5 stars).

Second, conversations lasted for on average one-fourth to one-third of an hour. Table~\ref{table:session_info} displays the mean, minimum, median, and maximum session lengths and time durations. The median session duration is 14.9 minutes, yet there are large variations and the longest session took 4 hours. While having relatively long session duration compared to a typical chat, the median number of utterances and words (contributed by both customers and agents) remained moderate at 28 and 381, respectively. This is because some utterances have long time gaps. For agents, this pause mostly corresponds to the time needed to check information after a quote ``Would you mind holding on for a few minutes while I check...'' Customers pause for various reasons, e.g., customers multi-tasking and returning to the chat screen infrequently. Long pauses are known to make conversations less cohesive and harder to follow~\cite{craig2014unlocking}. Hence, pauses in speaker's turn-taking is used as an input signal in predicting customer satisfaction in this research. 

{
\begin{table}[h]
  \caption{Session length information}
  \vspace{-0.3cm}
  \label{table:session_info}
  \begin{tabular}{ccccc}
    \toprule
    & Mean & Min & Median & Max \\ 
    \midrule
    Duration (min) & 19.0 & 0.1 & 14.9 & 280.9 \\ 
Utterance (number) & 35.3 & 4.0 & 28.0 & 585.0 \\ 
Words (number) & 469.6 & 6.0 & 381.0 & 13954.0 \\ 
  \bottomrule
\end{tabular}
\end{table}
}

\section{Research Methodology}

\subsection{Problem definition}

Identifying dissatisfied customers on a live chat service is a crucial objective in customer care. A customer with an unpleasant experience may no longer consider future purchases of the same service, but more importantly, they may engage in negative word of mouth by writing extremely negative reviews and leaving poor ratings on the Web~\cite{blodgett1995effects}. Because reviews establish social presences and emulate social norms, such negative feedback could have a detrimental impact on the retention of other customers~\cite{kumar2006research}. This is a major crisis that companies face in the age of social media~\cite{hennig2000relationship}. 
This research hence focuses on identifying ``dissatisfactory'' sessions and considers the following research question: given a small subset of live chat sessions with customer ratings and a larger set of unlabeled chat sessions, can we predict which sessions were likely judged as dissatisfactory by customers? 

Samsung's live chat data serve as excellent ground truth for our task. In particular, the survey results are immune to acquisition-led bias (i.e., ratings are positive since they are left by purchasers), because everyone who visits the chat service would already be using Samsung products. Furthermore, the ratings are not subject to social influence bias, because ratings are not shared across customers. This service, however, is not free from the under-reporting bias (i.e., those who are greatly satisfied or dissatisfied are more likely to rate)~\cite{goodwin1990consumer}. It could in fact be that extremely dissatisfied customers lost any intention to communicate further and left the chat service without taking the follow-up survey, as described in instances seen in service marketing:  ``\textit{Rather than seek redress, many of these dissatisfied consumers will instead exit.}''\cite{blodgett1995effects} This leads us to the investigation of the following hypothesis on the positivity bias:

\begin{quote}
{(H)}~\textit{Non-respondents of the live chat service are more likely to have dissatisfactory experiences than the respondents.}
\end{quote}	

Investigating the inherent bias in customer ratings that is described in the above hypothesis and devising methods to handle them is crucial for businesses. Therefore, we not only aim to build a classification model to identify dissatisfactory sessions, but also attempt to test the hypothesis on the positivity bias. To test this hypothesis, we need to be able to identify whether a customer was dissatisfied or not for each session with low error rates. Hence in this paper, we introduce new approaches to predict session dissatisfactions from live chat logs and compare their performances against state-of-the-art approaches.

We aggregate the 5-star survey responses into a dichotomous scale and group (i) Very Dissatisfied and Dissatisfied ratings as ``true'' votes representing dissatisfied customers in the prediction task and (ii) Average, Satisfied, and Very Satisfied ratings as ``false'' votes in the prediction task. The resulting dependent variable is a binary value of 1 or 0. From the live chat data set, both textual features, including the raw chat content, and non-textual features, including time gaps in speakers' turn-taking, were utilized.

\subsection{Prediction model}

Note that our task on predicting the overall customer satisfaction is different from the well-covered research problem, sentiment analysis~\cite{pang2008opinion}. While the goal of sentiment analysis is to identify affective states embodied in a given text, our task aims to predict customer satisfaction with which a session is likely to end. Affective states of a customer can vary as the conversation evolves, and hence one needs to consider the dynamic flow of chat conversations to understand customer satisfaction. Below we summarize existing key approaches that are suitable for the problem and introduce the deep-learning-based methods to predict customer dissatisfaction. 

\subsubsection{Existing approaches} 
Textual features of a dialog such as affective expressions are known as a key determinant of customer satisfaction in online business conversations. Based on previous studies~\cite{kim2010classifying,park2015mining,herzig2016predicting,althoff2016large}, textual features are more important than any other possible features (e.g., session length and disconnecting entity) for customer satisfaction. Here, two lines of approaches using textual features have been proposed: one is based on valence and the other on n-grams.

\begin{enumerate}
\item {Prediction with valence}: 
Prior studies built machine learning classifiers for conversation dialogs based on affective features~\cite{herzig2016predicting}. We implemented the random forest classifier that was proposed in the context of analyzing chat data~\cite{park2015mining}. Sentiment scores (i.e., positive or negative) in this classifier were extracted via VADER, a human-validated sentiment lexicon~\cite{hutto2014vader}, and then aggregated separately for agents and customers for every quarter of the session duration (i.e., 4 quarters). This resulted in a total of 8 affective features to be used in the classifier. We call this algorithm \textsf{Valence}.

\item {Prediction with $n$-grams}: 
$n$-grams include a contiguous sequence of $n$ items from text, where items can be either syllables, letters, or words. A common choice is a word token separated by white spaces. $n$-grams characterize the input text sequences and have been applied in prediction tasks of various domains, including live chat systems~\cite{kim2010classifying} and call centers~\cite{li2016customer}. In a recent study, $n$-grams have been used in predicting the success of counseling sessions for mental illness patients~\cite{althoff2016large}. A regression model was constructed with L1 regularization, and unigram and bigram features were found to be the most effective for prediction. In this paper, we implemented the same prediction model based on the top-1000 frequently appearing unigrams and bigrams. We call this algorithm \textsf{Ngram}.

\end{enumerate}

What is common in the above studies is the use of aggregated statistics such as the mean sentiment scores or n-gram frequencies from chat data. Efficiency of these feature-based approaches hence comes at the cost of information loss, in particular, on the exact linguistic choices (i.e., how word usages change from the beginning to the end of a dialog). To compensate for such information loss, existing studies segmented each chat dialog into different phases of conversation and repeatedly examined averaged statistics. Nonetheless, observations in aggregated data are limited by arbitrary time divisions, and important temporal dynamics remain missing. Another type of information loss that occurs due to aggregating data is the temporal evolution of chat responses, which is another important indicator of customer satisfaction. A study~\cite{zeithaml1988communication} found response promptness to be a critical factor in determining successful customer service. Information about response times to each utterance, when also aggregated at the level of the speaker or according to fixed chat duration, will be lost.  

The deep learning model utilized in this research effectively avoids any aggregation of data and the consequential loss of information that arises from the above feature-based approaches. The deep learning models suggested in this paper run over the sequence of text input rather than aggregated features, albeit at the cost of requiring heavy computing power.

\subsubsection{Deep-learning-based method}

Samsung's live chat sessions comprise multiple utterances, each of which is a sequence of word tokens. The recurrent neural network (RNN)-based models were trained with labeled data to learn the precise word usage patterns of dissatisfied customers and then applied to unlabeled sessions for prediction. The steps below first describe how to process the raw text input and then describe what structures are used in the RNN-based models.

\vspace*{2mm}
\noindent $\bullet$ \textbf{Preprocessing data}

\noindent
Live chats are dyadic conversations between customers and agents. From the raw chat data, we re-constructed multiple utterances as a sequence of word tokens representing each chat session, which is a desired input format of RNN-based models. Time gaps between two consecutive utterances were encoded along the word sequences in a similar manner to how previous studies handled time intervals in clickstreams~\cite{wang2016unsupervised}. We replaced every time gap with one of the followings: (1) \textit{{Short}$\_${Speaker}} to represent gaps less than the $25^{\text{th}}$ percentile of all such gap intervals from data, (2) \textit{{Medium}$\_${Speaker}} to represent gaps from the $25^{\text{th}}$ percentile to less than the $75^{\text{th}}$ percentile, and (3) \textit{{Long}$\_${Speaker}} to represent gaps of the $75^{\text{th}}$ percentile or longer between consecutive utterances. The suffix \textit{Speaker} indicates who determines how long a time gap will be---that is, the speaker who responds after the current utterance ends. 
From the chat logs, the $25^{\text{th}}$ percentile time gap for \textit{{Short}$\_${Customer}} was 13 seconds, indicating that a customer responded within 13 seconds after a given utterance by an agent. The \textit{{Long}$\_${Agent}} value was 49 seconds, indicating that an agent responded to a customer after a time gap of 49 seconds or longer. The percentile values on time gap distributions are presented in Table~\ref{table:timegap_info}.

{
\begin{table}[h]
  \caption{Percentile values on time gap distributions}
  \label{table:timegap_info}
  \begin{tabular}{ccccc}
    \toprule
    & $25^{\text{th}}$ & $50^{\text{th}}$ & $75^{\text{th}}$ & $100^{\text{th}}$ \\ 
    \midrule
    Agent & 12s & 26s & 49s & 439s \\ 
Customer & 13s & 27s & 51s & 1326s \\ 
  \bottomrule
\end{tabular}
\end{table}
}

\begin{figure}[t!]
{
\centering
\hspace{-8mm}
\includegraphics[width=\linewidth]{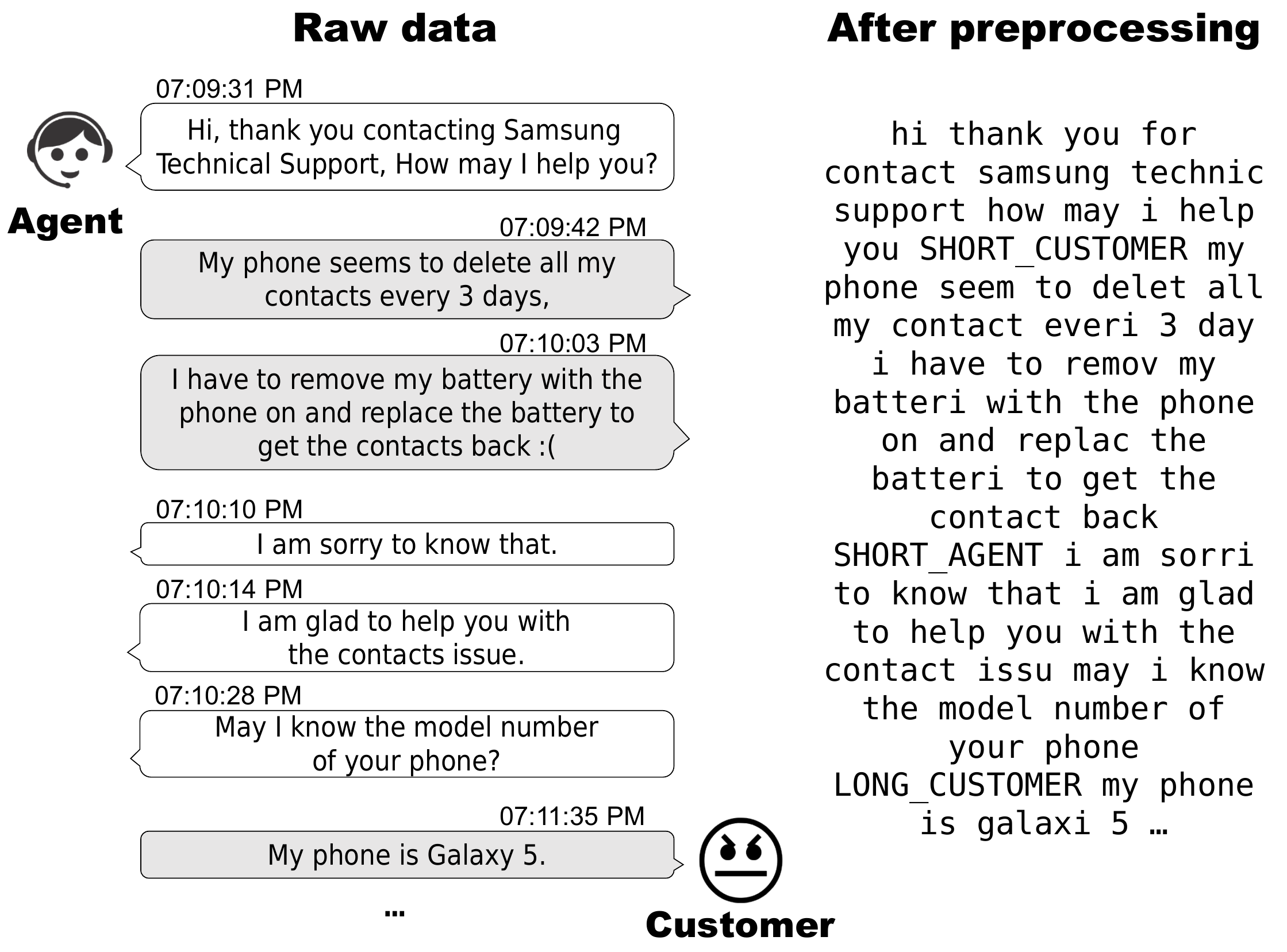}
 \caption{Illustration of preprocessing step}
 \label{fig:preprocessing}
}
\end{figure}

As mentioned earlier, we aggregated utterances at the level of speaker turn-taking and did not encode time gaps between utterances of the same speaker, for the reason that some speakers break a single sentence over multiple utterances by frequently pressing the enter key. Hence, a total of six identifiers were used to indicate time gaps in speaker turn-taking. Figure~\ref{fig:preprocessing} depicts how utterances of a sample session are transformed into a sequence of word tokens with time gaps. Following the practical guidelines from prior research~\cite{wang2016unsupervised}, one may interpret \textit{{Short}$\_${Speaker}} as representing engaged conversations, whereas \textit{{Medium}$\_${Speaker}} represents short pauses. \textit{{Long}$\_${Speaker}} may represent a long pause by agents or less engaged customers. 

After splitting utterances by white spaces into a list of word tokens, further preprocessing steps were followed, including stemming, filtering out special characters, and changing words to lower cases to reduce the complexity of word features based on the suggestions from previous works~\cite{tan2000textual,jones1997readings}. The less frequently appearing words were replaced with a special token.

\vspace*{2mm}
\noindent $\bullet$ \textbf{Model structure and training}

\noindent
A recurrent neural network is a type of artificial neural network that is designed to learn structures of any sequential representation of data such as text and voice. Recurrent neural networks possess a certain type of memory to preserve sequential information. Three kinds of structures are used to identify dissatisfied customers in this research. We briefly discuss the intuition behind the functioning of the three main models. 

\begin{enumerate}
\item The first model, called \textsf{RNN} in our evaluation, implements the standard neural network based on the most basic recurrent unit using a hyperbolic tangent nonlinearity function, $\tanh(\cdot)$. Here, an embedding layer is introduced to handle the sparsity of input sequences. The embedding layer transforms sparse word features into low-dimensional vector representations. Vectors produced by the embedding layer are fed into a hidden layer, which is composed of recurrent units. In the hidden layer, for each time step, a ${\tanh(\cdot)}$ function is used to update the current state by combining its previous state with the input of current state. Lastly, the binary label for customer dissatisfaction is predicted from the last step of the output layer. 

\item The second model is the \textsf{LSTM}, which stands for long short-term memory based neural network. Unlike the $\tanh$-based basic recurrent unit, an LSTM unit can remember the more distant past through its memory cell, called a gated cell~\cite{hochreiter1997long}. The gated cell makes efficient decisions about what to store and when to allow reads, writes and erasures via gates. Hence, this model is suitable for complex tasks such as time-series prediction~\cite{madetecting} and sequence modeling~\cite{tai2015improved}.
\item The third model is the \textsf{GRU} (gated recurrent unit) based on a variant of LSTM that has a simpler form. GRU does not have an output gate and, therefore, writes the full contents of its memory cell to the larger net at each time step. The simple structure of GRU makes it suitable for tasks involving a small amount of data, whereas LSTM requires a larger amount of data to train more parameters~\cite{chung2014empirical}. 
\end{enumerate} 
 
Except for the different recurrent units utilized in hidden layers (i.e., $\tanh$-based units, LSTM units, and GRUs), embedding layers were introduced in the same way across the three neural networks. A $\tanh$ function was used as the output squashing function. We trained all models by using the derivatives of the cross-entropy loss function via back-propagation through time. The Adam optimizer was used for parameter updates~\cite{kingma2014adam}. To prevent over-fitting, we applied dropout regularization~\cite{JMLR:v15:srivastava14a} to the hidden layer and L2 regularization to the last layer. We set the number of dimensions for embedding vectors as 50, each dropout rate as 0.2, the number of recurrent units as 500, and lambda for L2 regularizer as 0.001. These values were chosen via a grid search. Sequences were zero-padded when the length of a session is shorter than the number of recurrent units. The models were trained until the loss function converged in validation set (i.e., early stopping~\cite{prechelt1998automatic}) or the number of training epochs reached 100. The codes and implementation details are available on github\footnote{\url{https://github.com/bywords/Positivity-Bias-Livechat}}.

\section{Results}

\subsection{Evaluations}

Prior to performance evaluations, we first discuss the rationale for the binary label. One question arises on whether a star rating of 3 (Average) should be included in the true set or the false set. Should the Average star rating be more similar to Dissatisfied or Very Dissatisfied sessions than the rest, then its label must belong to the group of dissatisfied customers (i.e., the true set). This can be determined via measuring the distance between rater groups. Language vectors were constructed such that a global top-1000 unigram vector of measured word occurrences and term frequencies was produced for 10 percent of the sample sessions across labels. The cosine distance between the Average sessions and other sessions indicates that it is closer to Satisfied ($d$=0.186) and Very Satisfied ($d$=0.186) sessions than Dissatisfied ($d$=0.197) or Very Dissatisfied ($d$=0.219) sessions. For all cases, 95\% confidence interval ranges were smaller than 0.0007. We hence include the Average rating in the false set.

The final grouping contained 5,498 true sessions on dissatisfied customers and 21,559 false sessions on the remaining customers. These 27,057 sessions were randomly split into 80\% training set and 20\% test set. The training set was once more randomly split with a ratio of 80:20 to measure the loss function for validation purposes. The prediction model learning was performed on a balanced set of true and false instances, by randomly over-sampling each data set to avoid favoring popular sets. 

Table~\ref{table:prediction.res} displays the evaluation performances of the deep-learning-based models against two feature-based baseline models, \textsf{Valence} and \textsf{Ngram}. In particular, two versions of the deep learning models were implemented for \textsf{RNN}, \textsf{LSTM}, and \textsf{GRU}. The default version utilizes the chat content only and an extended version (appearing with suffix `-Time') also uses time gap information in chat utterances, as well as textual features. Precision and recall measure how precisely and with what sensitivity a given model predicts dissatisfied customers, respectively. F1 score is the harmonic mean of the two metrics and indicates a balanced score rather than accuracy. Thus, we mainly focus on F1 score for comparison.

We make the following observations. First, comparing the baseline models, we find that aggregated valence is not as effective as finer data structures such as $n$-gram. \textsf{Ngram} yielded a gain in F1 score of 0.24 over \textsf{Valence}. However, in regard to other metrics, \textsf{Valence} showed the highest precision of all, despite having only 8 simple linguistic features. This finding demonstrates the power of affective expressions in predicting customer satisfaction. Second, we find deep learning models show advantages over the feature-based baseline models. While \textsf{RNN} showed poor performance, both \textsf{LSTM} and \textsf{GRU} outperformed all alternatives. This finding implies that while it is hard to model long sequences through a simple RNN, which is known as the long-term dependency problem~\cite{bengio1994learning}, the other two models can look at far distant past within the chat conversation effectively. \textsf{GRU} excelled in achieving high precision, while \textsf{LSTM} overall showed the best performance in terms of F1 score due to its high recall. Third, we find that incorporating time gaps into \textsf{LSTM} (\textsf{LSTM-Time}) and \textsf{GRU} (\textsf{GRU-Time}) increased the F1 score by 0.0183 and 0.0171, respectively. This improvement shows that time gaps are a meaningful indicator of customer satisfaction. The performance of \textsf{RNN-Time}, however, degrades compared to that of \textsf{RNN}, possibly because this basic recurrent neural network has low capabilities in handling additional features and this lack of sophistication makes it more difficult to train temporal dynamics of word sequences.

{
\begin{table}[h]
	\frenchspacing
  \caption{Prediction results across 8 models with the top 2 values highlighted in bold text.}
  \label{table:prediction.res}
  \begin{tabular}{c|cccc}
    \toprule
    Method &Accuracy&Precision&Recall&F1\\ 
    \midrule
    \textsf{Valence} & 0.6416 & \textbf{0.8374} & 0.3516 & 0.4952\\
\textsf{Ngram} & 0.7668 & 0.7679 & 0.7054 & 0.7352\\ 
\midrule
\textsf{RNN} & 0.6912 & 0.6623 & 0.6683 & 0.6653\\
\textsf{LSTM} & 0.8005 & 0.7865 & \textbf{0.7764} & \textbf{0.7814}\\ 
\textsf{GRU} & 0.7984 & 0.8254 & 0.7116 & 0.7643\\ 
\textsf{RNN-Time} & 0.6609 & 0.6785 & 0.5078 & 0.5685\\
\textsf{LSTM-Time} & \textbf{0.8102} & 0.7758 & \textbf{0.8250} & \textbf{0.7997}\\ 
\textsf{GRU-Time} & \textbf{0.8106} & \textbf{0.8314} & 0.7371 & 0.7814\\
  \bottomrule
\end{tabular}
\end{table}
}

\subsection{Inferences on unlabeled sessions}

Having confirmed that deep learning methods can effectively classify dissatisfied customers using labeled data, we now turn our focus to the research hypothesis suggested in the previous section and infer which sessions likely contain dissatisfied customers based on unlabeled data. Before investigating how likely an unlabeled session is to be dissatisfactory, we used manual coding techniques to validate the prediction results. Three human coders participated in this step, who had experience handling the customer ratings data. First, the coders were provided with 50 randomly chosen labeled sessions to get familiarized with the chat data. Coders were provided with the full chat content as well as the meta-information, such as the session lengths and response times. Once the coders read through the labeled sessions, we then provided them with 100 randomly chosen unlabeled sessions for prediction. The coders were not told that these sessions were unlabeled, but were simply asked to label whether the customer of each session would be dissatisfied or not with the chat experience. Their responses were aggregated via majority voting. The tagging task showed high agreement rates of 0.508 based on unweighted Fleiss's Kappa (p$<$0.001). 

Treating the human-labeled data as ground truth, we compared the aggregated responses with labels predicted by \textsf{LSTM-Time} model. We used \textsf{LSTM-Time} for its consistently high F1 scores from experiments on the labeled dataset. The manually coded responses and the labels generated by our deep learning model had a moderate level of agreement rate in terms of unweighted Cohen's kappa ($\kappa$=0.296, p$<$0.01). Note that a value smaller than 0 indicates no agreement, $0$--$0.20$ is slight, and larger values indicate greater degrees of agreement. This finding shows that the prediction on the unlabeled dataset by \textsf{LSTM-Time} is similar to the truth, which increases the credibility of the inferred satisfaction from our approach. 

\begin{figure}[t]
{
\centering
\includegraphics[width=0.99\linewidth]{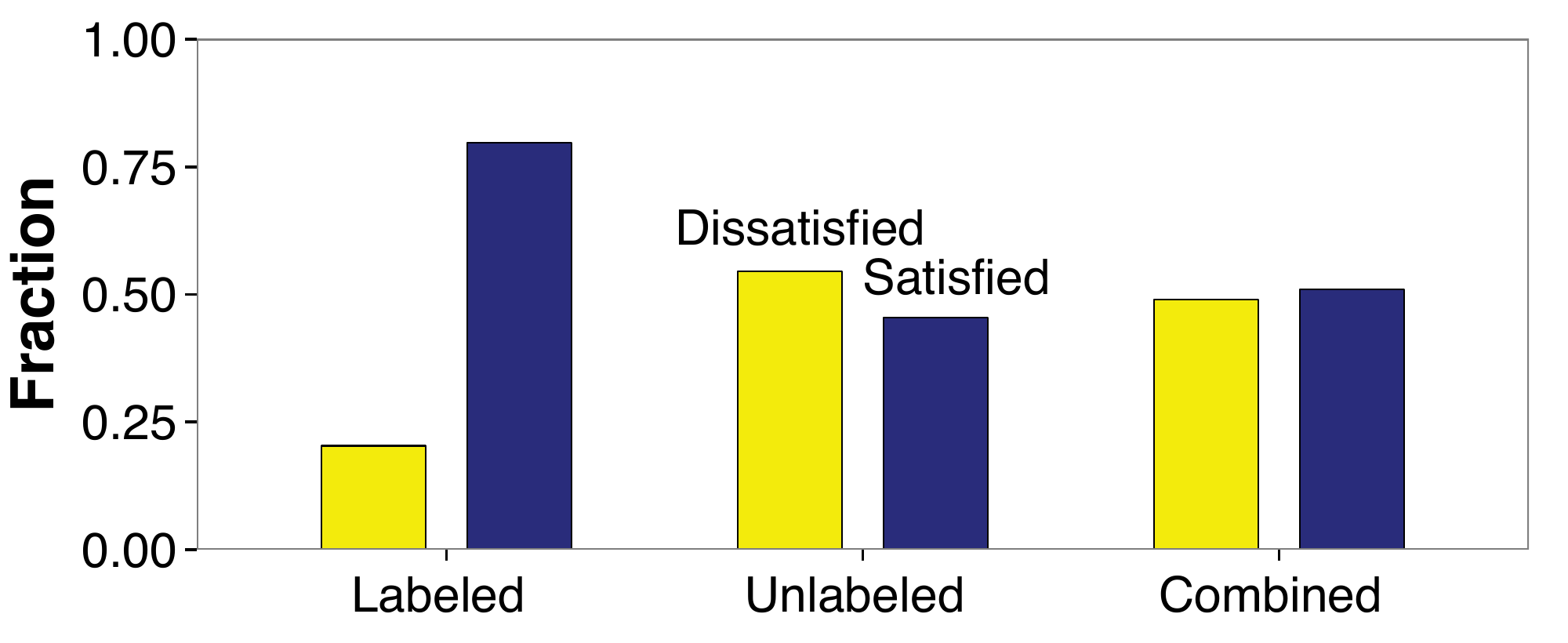}
 \caption{Fraction of dissatisfied sessions for labeled, unlabeled, and combined data}
 \label{fig:unlabel_prediction}
}
\end{figure}

Having validated the models on unlabeled data, finally we investigate distributions of customer satisfaction among raters and non-raters. Figure~\ref{fig:unlabel_prediction} compares the ratio of dissatisfactory sessions out of all sessions ({i}) from the labeled data with ground-truth, ({ii}) unlabeled data based on inference, and ({iii}) the two data sets combined. The results clearly demonstrate that non-raters are not as positive toward their chat experiences as the survey respondents. More than half of non-raters (54.5\%) are predicted to have been dissatisfied if they were to provide feedback. The chi-squared test confirms a meaningful difference between the labeled and unlabeled groups (${\chi}^2=$10623, p$<$0.001). This finding supports the main hypothesis of this paper that there exist positivity biases on Samsung's live chat service such that a signal of no feedback is more likely associated with customer dissatisfaction. Customer service centers hence need to incorporate the unlabeled sessions into their overall evaluations of customer satisfaction.

\section{Discussion \& Conclusion} 

Surveys have been extensively used to assess thoughts, opinions, and feelings of people in many different disciplines. While surveys have clear advantages, they often suffer from biases that hinder the generalization of findings to the target population. In the age of the Web and social media, passive online surveys such as customer ratings tend to entail extra biases such as social influence bias and under-reporting bias. Though many studies have qualitatively reported the existence of such distortion~\cite{mudambi2010makes}, little effort has been made to investigate and further correct them based on large data due to the difficulties in gathering data and designing analysis  methodologies~\cite{salganik2006experimental}. 

Gaining access to extensive and proprietary data describing chat logs and ratings, we had a unique opportunity to study customer satisfaction. The studied chat service is free from acquisition-led selection bias and social influence bias, yet it was under the influence of the positivity bias, to which under-reporting bias contributed. Toward investigating and compensating for this bias, we proposed deep learning approaches to infer ratings from data with high accuracy.

From the prediction on unlabeled sessions, empirical results reveal that the majority of non-respondents likely were dissatisfied with the chat service, unlike what had been reported by the survey respondents. The different natures of ratings for labeled and unlabeled sessions are evident in Figure~\ref{fig:unlabel_prediction}. When known scores and inferred scores are combined, reviews of the live chat service are no longer dominantly positive. Therefore, as similarly observed in previous studies on a different context~\cite{nosko2015limits,filippas2017reputation}, the findings of this research support the hypothesis on the positivity bias in rating systems.
We note that predictions for unlabeled sessions cannot be fully validated because such data are `unlabeled'. To address this limitation, human coders had been hired to obtain ground truth for a small set of unlabeled sessions, yet predictions for the unlabeled dataset still need further validation. Nonetheless, with the high performance of the proposed deep learning model and manual validation of predictions for the unlabeled dataset, our data analysis finds that customers who did not rate their experience with the chat system likely had more negative experiences. 

This finding is particularly important because past studies only utilized labeled data for investigating customer satisfaction and discarded the larger majority of unlabeled data in the analysis. It can be misleading to extrapolate general opinions of customers from survey responses, because respondents and non-respondents may possess different attitudinal traits~\cite{sax2003assessing}. This has practical implications because rated sessions are not representative of the overall customer opinion, and those unrated sessions need to be considered in conjunction with the rated ones to gain a full picture of online services such as live chat.

\subsection{Limitations and future work}

This research has several limitations. First, while the predictions for unlabeled sessions were validated through manual coding by three human coders, we only utilized a small fraction of the unlabeled sessions. Future research may rely on alternative methods like crowdsourcing on a larger dataset~\cite{wilson2016crowdsourcing}. Second, this work utilized standard RNN-based deep learning methods, limiting the length of observation. We plan to employ more sophisticated approaches to utilize all possible signals from a chat conversation such as attention models~\cite{luong2015effective} or convolutional neural networks~\cite{kim2014convolutional}. The last limitation is on the use of a single data source, due to the proprietary nature of data. 
Future studies could test the efficacy of the time gap features on multiple dataset. Moreover, other non-textual signals could be deeply investigated to understand customer satisfaction, and more broadly, to infer emotional states of speakers on a chat.

In the future, it would be meaningful to repeat this study for a wider range of service categories. For example, one can compare how ratings in closed proprietary systems compare to those in open systems that share ratings publicly. In addition, cultural norms may be another important factor that needs to be considered. While the studied live chat data set studied was mostly for customers in the United States, thousands of customers accessed from other parts of the world, including India, Canada, and the United Kingdom. Interestingly, survey response rates in these countries were significantly lower than in the US (1-3\% of all chat customers), despite showing similar satisfaction scores. Understanding how social norms affect online rating behaviors and what kind of biases prevail across cultures will help businesses and recommendation systems better utilize the customer ratings data. 

\section*{Acknowledgement}

We thank the anonymous reviewers for their constructive comments that further improved this paper. We also want to acknowledge the human coders for their contributions on manual annotations. This research was in part supported by the Next-Generation Information Computing Development Program (No. NRF-2017M3C4\\A7063570) and the Basic Science Research Program (No. NRF-2017R1\\E1A1A01076400) through the National Research Foundation (NRF) funded by the Ministry of Science and ICT of Korea.

{

}

\end{document}